\documentstyle[12pt,fleqn]{article}

\addtocounter{section}{-1}
\def\beeq{\begin{equation}}
\def\eneq{\end{equation}}
\def\beeqa{\begin{eqnarray}}
\def\eneqa{\end{eqnarray}}

\def\nel{N_{\rm el}}

\begin{document}

\vspace{0.8cm}
\begin{flushleft}
{\bf EXCITON EFFECTS AND NONLINEAR OPTICAL RESPONSE \\
IN SOLITON LATTICE STATES OF DOPED CONJUGATED POLYMERS}\\
\vspace{1.3cm}
KIKUO HARIGAYA, YUKIHIRO SHIMOI, AND SHUJI ABE\\
{\sl Electrotechnical Laboratory, Tsukuba, Ibaraki 305, Japan}
\end{flushleft}
\vspace{1.0cm}
\noindent
\underline{Abstract}
Exciton effects on conjugated polymers are investigated in
the soliton lattice system.  We use the Su-Schrieffer-Heeger
model with long-range Coulomb interactions treated by the
single-excitation configuration-interaction method.  The
soliton band is present in the Peierls gap of the doped system.
There appears a new kind of the exciton where an electron-hole
pair is excited between the soliton band and the continuum states.
We find that the oscillator strengths accumulate rapidly at this
exciton as the soliton concentration increases.  The contribution
from the lowest exciton is more than 90\% at the 10\% doping.
The third-harmonic generation (THG) at off-resonance frequencies
is calculated as functions of the soliton concentration and the
chain length of the polymer.  The optical nonlinearity by the
THG at the 10\% doping increases by the factor about 10$^2$ from
that of the neutral system.

\vspace{1.0cm}
\begin{flushleft}
\underline{INTRODUCTION}
\end{flushleft}

Recently, the roles and excitation structures of excitons in
conjugated polymers have been investigated intensively, relating
with the origins of nonlinear optical spectra.$^{1-4}$  The lowest
energy excitation has the largest oscillator strength as the
most remarkable consequence of correlation effects.  This feature
is observed when the correlations are taken into account
by the single-excitation configuration-interaction (single-CI)
method$^1$ and also by the time-dependent Hartree-Fock (HF) formalism.$^4$

It is widely known that the soliton lattice state is realized$^5$
when the Su-Schrieffer-Heeger (SSH) model$^6$ is doped with electrons
or holes.  The soliton band develops in the Peierls gap as the
doping proceeds.  When correlation effects are considered by the
single-CI, the excitation structures exhibit the presence of excitons.
In this paper, we consider exciton effects in the soliton lattice
state.  There is one kind of the exciton in the half-filled system,
where the excited electron (hole) sits at the bottom of the conduction
band (top of the valence band).  We call this exciton as the
``intercontinuum exciton".  In the soliton lattice state, there are
small gaps between the soliton band and the continuum states,
i.e., valence and conduction bands.  Therefore, the number of
the kind of excitons increases, and their presence is reflected
in structures of the optical spectra.  A new exciton, which we
name the ``soliton-continuum exciton", appears when the
electron-hole excitation is considered between the soliton
band and one of the continuum bands.  We look at variations of
relative oscillator strengths of the new excitons, the
soliton-continuum and intercontinuum excitons.

Next, we look at how the above changes of the characters of optical
excitations are reflected in the nonlinear optical properties.  We
will be able to realize large optical nonlinealities in the soliton
lattice system, because the energy gap is small in doped conjugated
polymers.  We consider the off-resonant nonlinear susceptibility
as a guideline of the magnitude of the nonlinearity.  We calculate
the third harmonic generation (THG) at zero frequency, with changing
the chain length and the soliton concentration.  We show that the
optical nonlinearity by the THG at the 10 percent doping increases
by the factor about 10$^2$ from that of the neutral system.  This is
owing to the accumulation of the oscillator strengths at the lowest
exciton with increasing the soliton concentration.

\begin{flushleft}
\underline{MODEL AND METHOD}
\end{flushleft}

The SSH hamiltonian$^6$ is considered with the long-range Coulomb
interactions, $1/\sqrt{(1/U)^2 + (r/a V)^2}$, whrere $U$
is the strength of the onsite interaction, $V$ means the strength
of the long range part, $r$ is the distance between atom sites,
and $a$ is the mean bond length.  The model is treated by the HF
approximation and the single-CI for the Coulomb
potential.  The adiabatic approximation is applied to the lattice.
The HF order parameters and dimerization amplitudes are determined
selfconsistently using the iteration method.  The details of the
formalism have been explained in the previous paper.$^7$  A geometry
of a ring with the radius $Na/2\pi$ is used for a polymer chain
with $N$ carbon sites.  The electric field of light is in the
molecular plane.  The THG is calculated with the sum-over-states
method.  For demonstration of the magnitude of the THG, we use the
value of the number density of the CH unit, which is taken from
{\sl trans}-polyacetylene: $N_{\rm d}=5.24\times 10^{22} {\rm cm}^{-3}$.$^8$
We also use the average hopping integral $t=1.8$eV in order to look
at numerical data in the esu unit.  The system size is chosen as
$N= 80, 100$, and 120, because the size around 100 is known to
give well the energy gap value of the infinite chain.  The excess
electron number is taken upto the 10\% doping.  We take one
combination of the Coulomb parameters $(U,V) = (4t,2t)$ as the
representative case.  The parameters in the SSH model$^6$ are
$t = 1.8$eV, the spring constant $K = 21$eV/\AA$^2$, and
the electron-lattice coupling $\alpha = 4.1$eV/\AA.
Most of the quantities of energy dimension are shown in the
units of $t$.

\begin{flushleft}
\underline{EXCITON EFFECTS IN OPTICAL SPECTRA}
\end{flushleft}

Figure 1(a) shows the optical absorption spectrum at the 2\%
soliton concentration.  The broadening $0.05t$ is used.
There are two main features around the energies 0.7$t$ and 1.4$t$.
The former originates from the soliton-continuum exciton, and
the latter is from the intercontinuum exciton.

Figure 1(b) displays the absolute value of the THG against the
excitation energy $\omega$.  The abscissa is scaled by the
factor 3 so that the features in the THG locate at the similar
points in the abscissa of Fig. 1(a).  The small feature at about
$\omega=0.22t$ comes from the lowest excitation of the
soliton-continuum exciton and the larger features at about
$\omega=0.24t$ and $0.32t$ come from the higher excitations.
The features from the intercontinuum exciton extend from
$\omega = 0.48t$ to the higher energies.  In the present
calculations, the THG in the energy region higher than 0.5$t$
is not large relatively.  The point, that the THG becomes
smaller as the excitation energy increases, has been seen
in the calculations of the half-filled system.$^4$  However,
in the time-dependent HF formalism, the THG in higher energies
is still larger as shown by Fig. 4 of ref. 4.  The
difference of the distribution of the THG strengths might
come from the difference of the approximation method for
electron correlations.

Figure 2 summarizes the optical gaps of the two kinds of
excitons.  The optical gap of the soliton-continuum exciton
is almost independent of the concentration.  The gap of
the intercontinuum exciton increases rapidly when the
concentration is larger than 2.5 percent.  This is due to
the fact that the number of states in the soliton band
increases, and thus the energy gap between continuum states
increases.

Figure 3 shows the ratio of the oscillator strengths of
the soliton-continuum exciton with respect to the total
oscillator strengths, plotted against the soliton
concentration.  The closed and open squares are the results
of the HF-CI and HF calculations, respectively.  The closed
squares have the larger ratio than the open ones.  This is
one of exciton effects.  When the concentration is near zero,
the ratio varies almost linearly.  This would be the natural
consequence for low concentrations, because interactions
among solitons are exponentially small and thus the ratio
is proportional to the number of solitons.  The increase
of the ratio saturates at about 7 percent.  The
soliton-continuum exciton becomes like a free exciton at
larger concentrations owing to the formation of the soliton band.

The THG data like in Fig. 1(b) are calculated for the three system
sizes, $N=80$, 100, 120, and for the soliton concentrations
up to 10\%.  As clearly seen for example in Fig. 1(b), the
off-resonant THG at $\omega = 0$ is quite far from features
coming from excitons.  The contributions from double (triple and so on)
excitations would be very small.  Thus, the single-CI calculations
could be used as a measure of the optical nonlinearities of
doped conjugated polymers.

Figure 4 displays the variations of the absolute
value of $\chi_{\rm THG}^{(3)} (0)$ for $(U,V) = (4t,2t)$.
The deviations of the plots from the expected smooth behavior
might come from the quantum effect due to the finite system size.$^9$
The linear absorption has the size consistensy, so the plots of the
relative oscillator strengths vary smoothly as functions of the
soliton concentration as shown in Fig. 3.  However, the THG is not
size consistent, and spectral shapes depend on the system size when
$N$ is as large as 100.$^9$  Therefore, it would not be
strange even if $|\chi_{\rm THG}^{(3)}| (0)$ is sensitive to
the system size and the soliton concentration.  The THG
increases as the system size increases.  This behavior is
the same as has been seen in the calculations of the half-filled
systems.$^9$  The increase of the off-resonant THG near zero
concentration is very rapid, but the THG is still increasing
for a few percent to 10\% soliton concentration.

Then, why such the large increase of the THG would occur upon
doping of the polymers?  In the soliton lattice theory by the
continuum model,$^5$ the energy gap decreases as the soliton
concentration increases.  Therefore, it may seem first that
the decrease of the energy gap is one of the reasons.  But,
as shown in the Fig. 2, the lowest optical gap is almost
independent of the concentration, and thus the change of the
optical gap would not be the main reason.  However, the ratio
of the oscillator strength of the soliton-continuum exciton increases
very rapidly.  In view of this change of the exciton characters,
it would be natural to conclude that the increase of the THG
by the factor 10$^2$ is due to the fact that the oscillator
strengths accumulate rapidly at the lowest exciton with increasing
the soliton concentration.

\begin{flushleft}
\underline{SUMMARY}
\end{flushleft}

Exciton effects have been investigated in the soliton lattice system.
There appears a new kind of the exciton where an electron-hole
pair is excited between the soliton band and the continuum states.
We have considered the off-resonant nonlinear susceptibility
as a guideline of the strength of the nonlinearity in the doped
conjugated polymers.   We have calculated the off-resonant
THG with changing the chain length and the soliton concentration.
We have shown that the optical nonlinearity at the 10 percent
doping increases by the factor about 10$^2$ from that of the
neutral system.  We have discussed that this is owing to the
accumulation of the oscillator strengths at the lowest exciton
with increasing the soliton concentration.

\begin{flushleft}
\underline{REFERENCES}
\end{flushleft}

\noindent
1. S. Abe, J. Yu, and W. P. Su, \underline{Phys. Rev. B},
\underline{45}, 8264 (1992).\\
2. S. Abe, M. Schreiber, W. P. Su, and J. Yu,
\underline{Phys. Rev. B}, \underline{45}, 9432 (1992).\\
3. D. Guo, S. Mazumdar, S. N. Dixit, F. Kajzar,
F. Jarka, Y. Kawabe, and N. Peyghambarian,
\underline{Phys. Rev. B}, \underline{48}, 1433 (1993).\\
4. A. Takahashi and S. Mukamel,  \underline{J. Chem. Phys.},
\underline{100}, 2366 (1994).\\
5. B. Horovitz, \underline{Phys. Rev. Lett.}, \underline{46}, 742 (1981).\\
6. W. P. Su, J. R. Schrieffer, and A. J. Heeger,
\underline{Phys. Rev. B}, \underline{22}, 2099 (1980).\\
7. K. Harigaya, Y. Shimoi, and S. Abe,
\underline{J. Phys.: Condens. Matter}, \underline{7}, 4061 (1995).\\
8. C. R. Fincher, C. E. Chen, A. J. Heeger, A.  G. MacDiarmid,
and J. B. Hastings, \underline{Phys. Rev. Lett.}, \underline{48},
100 (1982).\\
9. S. Abe, M. Schreiber, W. P. Su, and J. Yu, \underline{J. Lumin.}
\underline{53}, 519 (1992).\\

\mbox{}

\noindent
FIGURE 1 (a) The optical absorption spectrum and (b) the absolute
value of the THG, for the system size $N=100$, the electron
number $\nel = 102$, and $(U,V) = (4t,2t)$.  The broadening
$0.05t$ is used in (a), and $0.02t$ is used in (b).

\mbox{}

\noindent
FIGURE 2 The optical gaps in the single-CI of the ``intercontinuum
exciton" (filled squares) and of the ``soliton-continuum exciton"
(open squares), plotted against the soliton concentration.

\mbox{}

\noindent
FIGURE 3  The ratio of the total oscillator strength of the
``soliton-continuum exciton" as a function of the soliton concentration.
The open squares are the data for the HF absorption, while the filled
ones for the HF-CI absorption.

\mbox{}

\noindent
FIGURE 4  The absolute value of the THG at $\omega = 0$ v.s.
the soliton concentration for $(U,V) = (4t,2t)$, shown in the
esu unit.  The numerical data are shown by the triangles ($N=80$),
circles ($N=100$), and squares ($N=120$), respectively.
The dashed lines are the guide for eyes.

\end{document}